\documentclass[prl,aps,graphics,epsfig,twocolumn,floats,superscriptaddress,showpacs]{revtex4}
\hfuzz=\maxdimen
\usepackage{graphicx}
\usepackage{amssymb}
\usepackage{amsmath}
\usepackage{color}

\begin{document}

\title{Galvanomagnetic effects and manipulation of antiferromagnetic interfacial uncompensated magnetic moment in exchange-biased bilayers}

 \author{X. Zhou}
\affiliation{Shanghai Key Laboratory of Special Artificial Microstructure and Pohl Institute of Solid State
Physics and School of Physics Science and Engineering, Tongji University, Shanghai 200092, China}
\author{L. Ma}
\affiliation{Shanghai Key Laboratory of Special Artificial Microstructure and Pohl Institute of Solid State
Physics and School of Physics Science and Engineering, Tongji University, Shanghai 200092, China}
\author{Z. Shi}
\affiliation{Shanghai Key Laboratory of Special Artificial Microstructure and Pohl Institute of Solid State
Physics and School of Physics Science and Engineering, Tongji University, Shanghai 200092, China}
\author{W. J. Fan}
\affiliation{Shanghai Key Laboratory of Special Artificial Microstructure and Pohl Institute of Solid State
Physics and School of Physics Science and Engineering, Tongji University, Shanghai 200092, China}
\author{R. F. L. Evans}
\affiliation{Department of Physics, University of York, York YO10 5DD, United Kingdom}
\author{R. W. Chantrell}
\affiliation{Department of Physics, University of York, York YO10 5DD, United Kingdom}
\author{S. Mangin}
\affiliation{Institut Jean Lamour, UMR CNRS 7198, Universit¨¦ de Lorraine- boulevard des aiguillettes, BP 70239, Vandoeuvre cedex F-54506, France}
\author{H. W. Zhang}
\affiliation{
State Key Laboratory of Electronic Thin Films and Integrated Devices, University of Electronic Science and Technology of China, Chengdu 610054, China}

 \author{S. M. Zhou$^{\ddag}$ \footnotetext{${}^{\ddag}$ Correspondence author. Electronic mail: shiming@tongji.edu.cn}}
\affiliation{Shanghai Key Laboratory of Special Artificial Microstructure and Pohl Institute of Solid State
Physics and School of Physics Science and Engineering, Tongji University, Shanghai 200092, China}

\date{\today}
\vspace{5cm}

\begin{abstract}
In this work, IrMn$_{3}$/insulating-Y$_{3}$Fe$_{5}$O$_{12}$ exchange-biased bilayers are studied. The behavior of the net magnetic moment $\Delta m_{AFM}$ in the antiferromagnet is directly probed by anomalous and planar Hall effects, and anisotropic magnetoresistance. The $\Delta m_{AFM}$ is proved to come from the interfacial uncompensated magnetic moment. We demonstrate that the exchange bias and rotational hysteresis are induced by the irreversible switching of the $\Delta m_{AFM}$. In the training effect, the $\Delta m_{AFM}$ changes continuously. This work highlights the fundamental role of the $\Delta m_{AFM}$ in the exchange bias and facilitates the manipulation of antiferromagnetic spintronic devices.

\end{abstract}

\pacs{75.30.Gw; 75.50.Ee; 75.47.-m; 75.70.-i}

\maketitle

\indent Exchange bias (EB) phenomenon in ferromagnetic (FM)/antiferromagnetic (AFM) systems has attracted lots of attention because of its intriguing physics and technological importance in spin valve based magnetic devices~\cite{1Nogues1999,2Berkowitz1999,2aOgrady2010,4Dieny1991}. After the FM/AFM bilayers are cooled under an external magnetic field from high temperatures to below the N\'{e}el temperature of the AFM layers, the hysteresis loops are simultaneously shifted and broadened~\cite{5Meiklejohn1956}. FM/AFM bilayers are now commonly integrated in spintronic devices~\cite{6aPark2011}. Nevertheless manipulation and characterization of the AFM spins are important to understand and control the exchange bias phenomenon~\cite{52Fan2014}.\\
\indent Rotatable and frozen AFM spins are generally thought to be responsible for the coercivity enhancement and shift of the FM hysteresis loops~\cite{7Stiles1999,8Wu2010,10Ge2013,11Geshev1999,Evans2011}. Ohldag~\textit{et al} found that a nonzero AFM net magnetic moment $\Delta m_{AFM}$ is necessary to establish the EB~\cite{16Ohldag2003}. However, Wu \textit{et al} thought that the EB can be established without frozen AFM spins~\cite{8Wu2010}. Therefore, the behavior of AFM spins is still under debate.
Moreover, the EB training effect is attributed to the relaxation of the $\Delta m_{AFM}$ towards the equilibrium state during consecutive hysteresis loops~\cite{26Hoffmann2004,47Brems2005,28Biternas2010,29Su2012,33Zhang2001,34Qiu2008}. For FM/AFM bilayers, the rotational hysteresis loss at $H$  larger than the saturation magnetic field is ascribed to the irreversible switching of AFM spins during clock wise (CW) and counter clock wise (CCW) rotations~\cite{38Stiles1999,39Beckmann2003,40Tsunada2007,41Gao2007}. Since there is still a lack of direct experimental evidence, it is necessary to elucidate the fundamental mechanism of the AFM spins in FM/AFM bilayers in experiments.\\
\begin{figure}[tb]
\begin{center}
\resizebox*{4 cm}{4 cm}{\includegraphics*{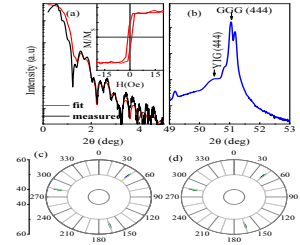}}
\caption{(a)Small angle x-ray reflection, (b)large angle x-ray diffraction on IrMn/YIG films, $\Phi$ and $\Psi$ scan with fixed $2\theta$ for the (008) reflection of GGG substrate (c) and YIG film (d). The room temperature in-plane magnetization hysteresis loop of the YIG layer is shown in the inset of (a). } \label{Fig1}
\end{center}
\end{figure}
\indent In most studies, the information of AFM spins is \textit{indirectly} explored from the hysteresis loops of the FM layers with micromagnetic simulations and Monte Carlo calculations~\cite{Evans2011,28Biternas2010,39Beckmann2003,47Brems2005}. In sharp contrast, very few methods can be implemented to directly probe the AFM spins due to almost zero net magnetic moment of the AFM layers. However, different measurements, combining x-ray magnetic circular dichroism and x-ray magnetic linear dichroism can detect FM and AFM spins due to their element-specific advantage~\cite{8Wu2010,13Ohldag2001,16Ohldag2003}. Only very recently, tunneling anisotropic magnetoresistance (TAMR) effect, which was initially proposed for tunneling device consisting of a single FM electrode and a nonmagnetic electrode~\cite{19Gould2004}, has been used to probe the motion of the AFM spins in AFM spintronic devices~\cite{17Marti2012,18Wang2012}. Since the TAMR arises from the tunneling density of states which depends on the orientation of the AFM spins in a \textit{complex} way, however, the orientation of the AFM spins cannot be determined directly and in particular the issue whether the $\Delta m_{AFM}$ does exist or not is still unsolved~\cite{20Shick2006}. In this Letter, we demonstrate clear evidence of the $\Delta m_{AFM}$ and reveal its mechanism behind the EB, the training effect, and the rotational hysteresis for IrMn$_{3}$(=IrMn)/Y$_{3}$Fe$_{5}$O$_{12}$ (YIG) bilayers using anomalous Hall effect (AHE), planar Hall effect (PHE), and anisotropic magnetoresistance (AMR) measurements. Here, the YIG \textit{insulator} is used as the FM layer such that all magnetotransport properties are contributed by the metallic IrMn layer. Galvanomagnetic measurements allow to probe the entire IrMn layer and not only the interface as in the reported TAMR measurements~\cite{17Marti2012,18Wang2012}. The $\Delta m_{AFM}$ in metallic IrMn is proved experimentally to arise from the interfacial uncompensated magnetic moment. It is clearly demonstrated in experiments that the EB and related phenomena are intrinsically linked to the partial pinning and irreversible motion of the $\Delta m_{AFM}$.\\
\begin{figure}[tb]
\begin{center}
\resizebox*{5 cm}{5 cm}{\includegraphics*{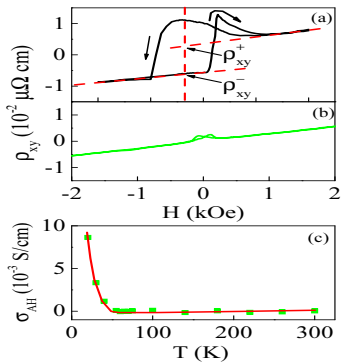}}
\caption{For IrMn/YIG bilayer, Hall loops at 20 K (a) and 50 K (b) with $H$ along the film normal direction, and AHC as a function of $T$ (c).} \label{Fig2}
\end{center}
\end{figure}
\indent IrMn (5 nm)/YIG (20 nm) bilayers were fabricated by pulsed laser deposition (PLD) and subsequent magnetron sputtering in ultrahigh vacuum on (111)-oriented, single crystalline Gd$_{3}$Ga$_{5}$O$_{12}$ (GGG) substrates~\cite{supple}. X-ray reflectivity (XRR) measurements show that YIG and IrMn layers are $20\pm 0.6$ and $5.0\pm0.5$ nm, respectively, as shown in Fig.~\ref{Fig1}(a).
The root mean square surface roughness of the YIG layer is fitted to be 0.6 nm. The x-ray diffraction (XRD) spectra in Fig.~\ref{Fig1}(b) show that the GGG substrate and YIG film are of (444) and (888) orientations. The pole figures in Figs.~\ref{Fig1}(c) and~\ref{Fig1}(d) confirm the epitaxial growth of the YIG film. As shown in inset of Fig.~\ref{Fig1}(a), the magnetization (134 emu/cm$^{3}$) of the YIG film is close to the theoretical value of 131 emu/cm$^{3}$ and the coercivity is small as 6 Oe. \\
\indent Before measurements, the films are patterned into normal Hall bar and then cooled from room temperature to 5 K under $H=30$ kOe along the film normal direction. The Hall resistivity $\rho_{xy}$ was measured as a function of the out-of-plane $H$ at various temperatures, as shown in Fig.~\ref{Fig2}. The Hall resistivity at spontaneous states $\rho_{xy}^{+}$ and $\rho_{xy}^{-}$ were extrapolated from the positive and negative high $H$ and the anomalous Hall resistivity was obtained by the equation $\rho_{AH}$ =$(\rho_{xy}^{+}-\rho_{xy}^{-})/2$. One has the anomalous Hall conductivity (AHC) $\sigma_{AH}\approx \rho_{AH}$/$\rho_{xx}^2$ because $\rho_{AH}$ is two orders of magnitude smaller than the $\rho_{xx}$~\cite{22Nagaosa2010}. Since the $\rho_{AH}$ decreases sharply and vanishes near $T=50$ K as shown in Figs.~\ref{Fig2}(a) and~\ref{Fig2}(b),
 the $\sigma_{AH}$ is reduced with increasing $T$ and is equal to zero at $T\geq50$ K in Fig.~\ref{Fig2}(c). More remarkably, since the shifting and asymmetry of the Hall loop both exist at low $T$ and vanish near the same $T=50$ K, the AHC is accompanied by the perpendicular EB~\cite{24Maat2001,24aZhou2004}. \\
\indent It is essential to address the physics for the AHC in the IrMn/YIG bilayers. With large atomic spin-orbit coupling of  heavy Ir atoms and magnetic moment (2.91 $\mu_B$) of Mn atoms, reasonably large AHC is expected in the chemically-ordered $L1_{2}$ IrMn alloy under high $H$~\cite{21Chen2014}, and it should be independent of the film thickness and change slowly with $T$ due to the high N\'{e}el temperature of the AFM alloy. In sharp contrast, the $\sigma_{AH}$ in the present IrMn/YIG bilayers changes strongly with the IrMn layer thickness, demonstrating an interfacial nature, as shown in Fig.S1~\cite{supple}. Therefore, the present AHC results should not be caused by the noncollinear spin structure on the kagome lattice~\cite{21Chen2014}, which is further confirmed by the vanishing AHC for the 5 nm thick IrMn films on GGG substrates in Fig.S2~\cite{supple}. This is because the present IrMn layers deposited at the ambient temperature are of the chemically disordered face-centered-cubic structure~\cite{Kohn2013}. With the strong $T$ dependence, the present AHC results cannot be attributed to the spin Hall magnetoresistance either~\cite{22aNakayama2013}. As pointed above, however, the AHC is strongly related to the established EB. As shown by the AMR results below, any FM layer at the interface can be excluded. Therefore, the AHC exclusively hints the existence of the IrMn interfacial uncompensated magnetic moment which is produced by the field cooling procedure. \\
\begin{figure}[tb]
\begin{center}
\resizebox*{5 cm}{!}{\includegraphics*[angle=90]{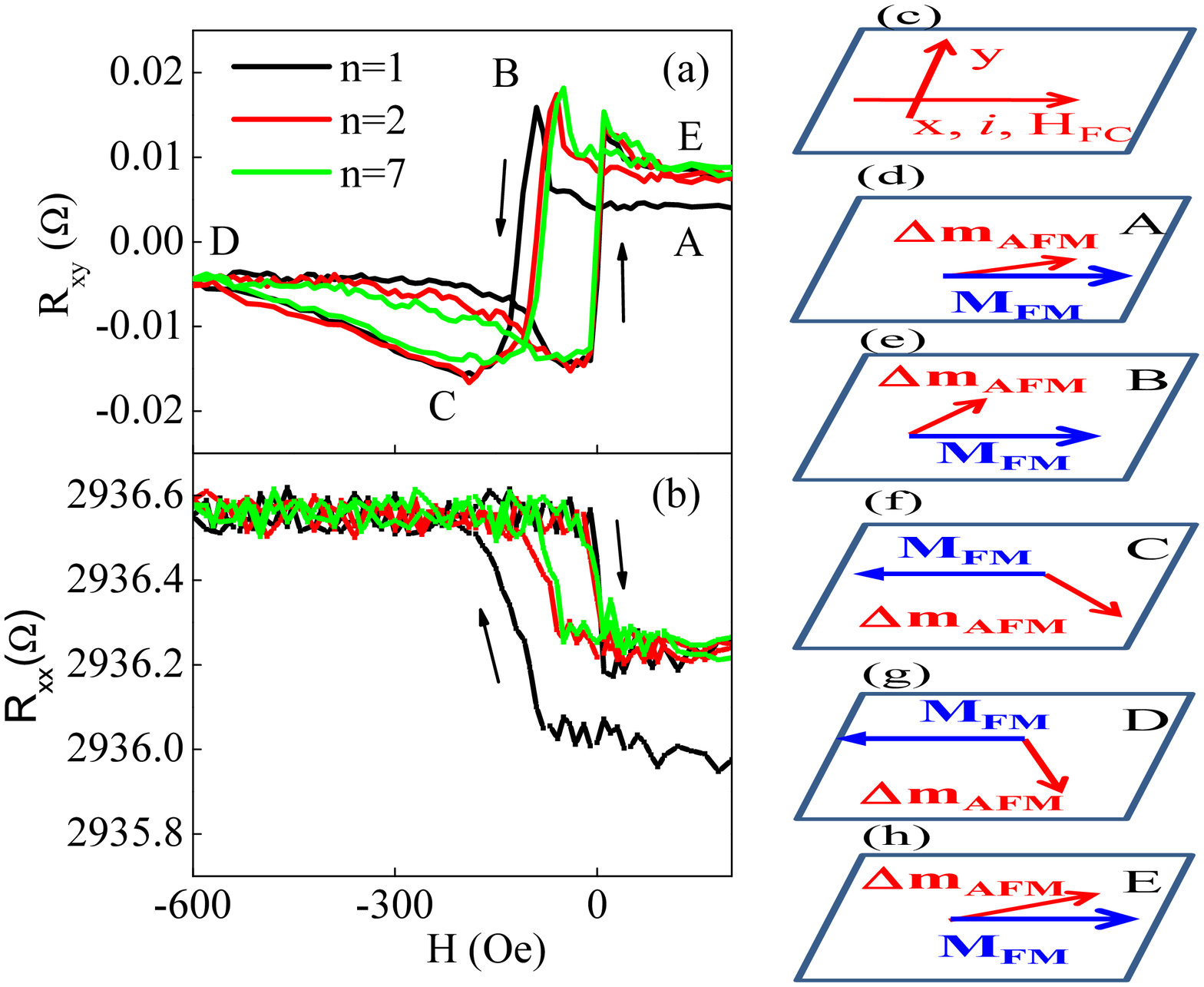}}
\caption{PHE loops (a) and AMR curves (b) of the cycle $n=1$, 2, and 7 at 5 K. In the schematic picture (c), the film is aligned in the \textit{x-y} plane, the sensing current \textit{i}, cooling field $H_{FC}$, and $H$ are parallel to the \textit{x} axis. The orientations of the $\Delta m_{AFM}$ and the FM magnetization are given at stages A(d), B(e), C(f), D(g), and E(h) of the descent branch of the $n=1$ in (a). In (d, e, f, g, h), $0<\theta_{AFM}(A)<\theta_{AFM}(B)<90^{0}$, and $-90^{0}<\theta_{AFM}(D)<\theta_{AFM}(C)<-\theta_{AFM}(B)$, and $0<\theta_{AFM}(A)<\theta_{AFM}(E)<90^{0}$. } \label{Fig3}
\end{center}
\end{figure}
\indent Figures~\ref{Fig3}(a) and \ref{Fig3}(b) show the PHE loops and AMR curves with consecutive cycles after the sample is cooled from room temperature to 5 K with the in-plane $H$ along the cooling field $H_{FC}$ which is parallel to the sensing current $i$, as shown in Fig.~\ref{Fig3}(c). Several distinguished features are demonstrated in the descent branch of the first cycle, $n=1$. Most importantly, for FM  metallic films the AMR curves are symmetric, that is to say, the values of the $R_{xx}$ at positive and negative high $H$ are equal to each other~\cite{25aMcguire1975,47Brems2005}. In striking contrast, however, the AMR curve in Fig.~\ref{Fig3}(b) is asymmetric. Therefore, the present AMR results cannot be attributed to any metallic FM layer at the interface but exclusively to the interfacial uncompensated magnetic moment of the IrMn layer. Accordingly, the PHE signal and the AMR ratio are proportional to $sin(2\theta_{AFM})$ and $1-cos^2\theta_{AFM}$, respectively, where $\theta_{AFM}$ refers to the orientation of the $\Delta m_{AFM}$ with respect to the $x$ axis~\cite{25aMcguire1975}. More remarkably, with the monotonic change of the $R_{xx}$, one has $|\theta_{AFM}|\leq90$ degrees. In combination with the sign change of the $R_{xy}$, the $\Delta m_{AFM}$ should be in either the first or the fourth quadrant~\cite{17Marti2012}, as schematically shown in Figs.~\ref{Fig3}(d)-\ref{Fig3}(g). The IrMn layer is \textit{far} from the negative saturation within the field of -600 Oe. Therefore, the angle between the FM and AFM spins is smaller (larger) than 90 degrees at the positive (negative) high $H$, and the FM/AFM system is of low (high) interfacial exchange coupling energy, leading to the lateral and vertical shift of the hysteresis loops~\cite{16Ohldag2003,25bLiu2004,25czhou}. Moreover, when the $H$ changes from stages B to C, the $\Delta m_{AFM}$ is irreversibly switched from the first quadrant to the fourth one~\cite{39Beckmann2003}, as demonstrated by the variations of $R_{xx}$ and $R_{xy}$ in Figs.~\ref{Fig3}(a) and ~\ref{Fig3}(b). As schematically shown in Figs.~\ref{Fig3}(d)-~\ref{Fig3}(g), during the sweeping of $H$, both the magnitude and orientation of the $\Delta m_{AFM}$ may change, indicating the multidomain process. Therefore, the observations of both the $\Delta m_{AFM}$ and its motion help to elucidate the intriguing physics behind the shifting and broadening of the hysteresis loops in FM/AFM bilayers, and in particular asymmetric magnetization reversal process of the FM magnetization~\cite{18Wang2012,25Zhou2004}. \\
\indent For the cycle number $n=1$, 2, and 7, the descent branch shifts significantly whereas the ascent branch almost does not change as shown in Figs.~\ref{Fig3}(a) and~\ref{Fig3}(b), in agreement with the first kind of the EB training effect of the (FM) magnetization hysteresis loops in Fig.S3~\cite{supple,33Zhang2001}. The athermal training effect from $n=1$ to $n=2$ is much larger than those of $n>2$, which was explained as a result of the switching of AFM spins among easy axes by Hoffmann~\cite{26Hoffmann2004}. In particular, the PHE signal and AMR ratio at the starting stage A are smaller than those of the ending stage E, indicating that the state of the $\Delta m_{AFM}$ \textit{cannot} be recovered after the first cycle, which further confirms the theoretical predictions~\cite{26Hoffmann2004,28Biternas2010,29Su2012}. As schematically shown in Figs.~\ref{Fig3}(a) and \ref{Fig3}(b), the $\Delta m_{AFM}$ experiences different trajectories during consecutive cycles, explaining the physics behind the EB training in FM/AFM systems~\cite{33Zhang2001,26Hoffmann2004,47Brems2005,41Gao2007,34Qiu2008,46wang2002}.\\
\indent Figures~\ref{Fig4}(a)-\ref{Fig4}(d) show the PHE signal as a function of $\theta_H$ with CW and CCW rotations under different magnitudes of $H$. At $H=50$ Oe, the CW and CCW curves overlap and the FM and AFM spins are expected to rotate reversibly within a small angular region. For higher $H$, the hysteretic behavior begins to occur and becomes strong for $H=300$ and 500 (Oe). This effect starts to decrease for $H=1.0$ kOe but still persists at $H=20$ kOe. Figures~\ref{Fig4}(e)-~\ref{Fig4}(h) show the angular dependence of the PHE signal with CW and CCW rotations under $H=1.0$ kOe at different $T$. At low $T$, there is a difference between the CW and CCW curves, indicating irreversible rotation of the $\Delta m_{AFM}$, and the hysteretic effect becomes weak at enhanced $T$. Near $T_B$, the measured results can be fitted well with $sin(2\theta_H)$ due to the ordinary magnetoresistance effect, as demonstrate by the vanishing PHE signal in Fig.S4~\cite{supple}. In a word, the hysteretic behavior of the PHE curves reproduce the rotational hysteresis loss of the FM magnetization in FM/AFM bilayers~\cite{1Nogues1999,5Meiklejohn1956,40Tsunada2007}.\\
\begin{figure}[tb]
\begin{center}
\resizebox*{5 cm}{!}{\includegraphics*{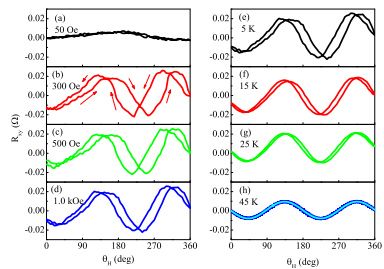}}
\caption{Angular dependent PHE signal with CW and CCW senses at $H=50$ (a), 300 (b), 500 (c), and 1000 (d) (Oe) , and at $T=5$ (e), 15 (f), 25 (g), and 50 (h) (K). $T=5$ K in the left column and $H=1.0$ kOe in the right column. In (h), solid cyan line refers to the $sin(2\theta_H)$ fitted results.} \label{Fig4}
\end{center}
\end{figure}
\indent It is interesting to analyze the magnitude and reversal mechanism of the $\Delta m_{AFM}$ as a function of $T$. The galvanomagnetic effects in Fig.~\ref{Fig2} and Fig.S4\cite{supple} become weak with increasing $T$, clearly indicating that the
 $\Delta m_{AFM}$ is reduced at elevated $T$ and approaches vanishing at $T_B$. Meanwhile, the $\Delta m_{AFM}$ at low $T$ is reversed irreversibly, leading to the EB establishment. At high $T$, reversible reversal becomes dominant, resulting in the disappearance of the EB. The variation of the reversal mode with $T$ confirms the validity of the thermal fluctuation model for \textit{polycrystalline} AFM systems~\cite{43Fulcomer1972,7Stiles1999,41Gao2007}. In this model, the reversal possibility is governed by the Arrhenius-N\'{e}el law and determined by the competition between the thermal energy and the energy barrier which is equal to the product of the uniaxial anisotropy and the AFM grain volume. The low $T_B$ of 50 K is induced by ultrathin thickness and the microstructural deterioration of the IrMn layer which is induced by the lattice mismatch between IrMn and YIG layers~\cite{17Marti2012}. At $T<T_B$, the energy barrier is larger than the thermal energy, leading to the irreversible process in most AFM grains. Accordingly, the EB is established and accompanied by the sizeable galvanomagnetic effects. Since more AFM grains become superparamagnetic for $T$ close to $T_B$, the $\Delta m_{AFM}$, galvanomagnetic effects, and the EB all approach vanishing, as shown in Figs.~\ref{Fig2} and~\ref{Fig3}, and Fig.S4\cite{supple}. On the other hand, the Meiklejohn-Bean model and the domain state model are \textit{not} suitable for the present results~\cite{5Meiklejohn1956,51Nowak2002}. In the former model, AFM spins are fixed during the reversal of the FM magnetization, which is in contradiction with the present results. In the latter one, the uncompensated AFM magnetic moment is mainly contributed by the bulk AFM~\cite{51Nowak2002} whereas the $\Delta m_{AFM}$ in the present IrMn/YIG systems mainly stems from the uncompensated magnetic moment at FM/AFM interface. \\
\indent In summary, for IrMn/YIG bilayers the interfacial uncompensated magnetic moment $\Delta m_{AFM}$ is observed by the galvanomagnetic effects. The partial pinning and irreversible switching of the $\Delta m_{AFM}$ are directly proved to be the physical source for the exchange field, coercivity enhancement, and the rotational hysteresis loss. The orientation of the $\Delta m_{AFM}$ is found to continuously change during the EB training effect. The present work permits a better understanding of the EB and related phenomena in FM/AFM bilayers. It demonstrates that galvanomagnetic measurements allow to probe the behavior of the AFM layer and consequently are a powerful tool to understand FM/AFM systems. This technique should be useful in the field of AFM spintronics. \\
\indent This work was supported by the State Key Project of Fundamental Research Grant No. 2015CB921501, the National Science Foundation of China Grant Nos.11374227, 51331004, 51171129, and 51201114, Shanghai Science and Technology Committee Nos.0252nm004, 13XD1403700, and 13520722700. \\

\end{document}